\documentclass[leqno,draft,11pt]{article}
\usepackage{amssymb,amsmath,latexsym,theorem,a4wide}

\def\sameenum{}

\ifx\mydefs\undefined\else \fi
\let\mydefs\relax

\ifx\slide\undefined
  \allowdisplaybreaks[2]
\fi

\ifx\loadcyr\undefined\else
\input cyracc.def
\makeatletter
 at 1\@ptsize pt
 at 1\@ptsize pt
 at 1\@ptsize pt
\makeatother

\fi

\def\gobble#1{}
\def\fixsup#1#2{{#1\let\dp\gobble\mathstrut}^#2_}

\def\bme{\hskip.75em\relax}

\let\eq\leftrightarrow
\let\EQ\Leftrightarrow

\def\iff{\quad\text{iff}\quad}
\let\ET\bigwedge
\let\TO\Rightarrow
\def\?{\mathbin?}

\newbox\circlebox
\setbox\circlebox\hbox{$\bigcirc$}
\def\circled#1{%
  \setbox0\hbox to\wd\circlebox{\hss$#1$\hss}\wd0=0pt
  \box0\copy\circlebox}

\let\fii\varphi
\let\tet\vartheta
\let\ep\varepsilon

\ifx\slide\empty
  \def\greek#1{$\expandafter\greeknum\csname c@#1\endcsname$}
\else
  \def\greek#1{$\mathop{\boldsymbol{\expandafter\greeknum\csname c@#1\endcsname}}$}
\fi
\makeatletter
\def\greeknum#1{\ifcase#1\or\alpha\or\beta\or\gamma\or\delta\or\ep
      \or\digamma\or\zeta\or\eta\or\tet\or\iota\else\@ctrerr\fi}
\makeatother

\def\p#1{\langle#1\rangle}

\let\sset\subseteq

\let\onto\twoheadrightarrow

\newcommand\rpair[3][3em]{\mathrel{%
   \begin{matrix}%
     \strut\smash{\xrightonto{\hbox to#1{\hss$#2$\hss}}}\\[-1.7ex]%
     \strut\smash{\xleftembed[\hbox to#1{\hss$#3$\hss}]{}}%
   \end{matrix}}}
\makeatletter
\newcommand\xrightonto[2][]{\ext@arrow 0359\rightontofill{#1}{#2}}
\newcommand\xleftembed[2][]{\ext@arrow 3095\leftembedfill{#1}{#2}}
\def\leftembedfill{\arrowfill@\leftarrow\relbar\hookleftnoarrow}
\def\rightontofill{\arrowfill@\relbar\relbar\onto}
\def\hookleftnoarrow{\DOTSB\relbar\joinrel\rhook}
\makeatother

\mathchardef\#="2023 

\ifx\busspf\undefined\else
\usepackage{bussproofs}
\EnableBpAbbreviations

\fi

\ifx\symlasy\undefined

  \def\centerdot#1{{%
    \setbox0\hbox{$\mathop{#1}$}\dimen0 \ht0
    \setbox0\hbox{$#1$}\advance\dimen0 -\ht0
    \setbox2\hbox to\wd0{\hss$\mathop{\cdot}$\hss}\wd2=0pt
    \lower\dimen0\box2\box0 }}
\else
  
  \def\centerdot#1{{%
     \setbox0\hbox{$#1$}%
     \raise0.206\ht0\hbox to\wd0{\hss$\cdot$\hss}%
     \kern-\wd0 \box0 }}
\fi

\def\ru{\mathrel/}

\let\sls|

\def\Up{{\setbox0\hbox{$\uparrow$}%
         \lower\dp0\hbox to\wd0{\hss\vrule width4pt height.4pt\hss}%
         \kern-\wd0\box0}}
\def\UP{{\setbox0\hbox{$\uparrow$}%
         \lower\dp0\hbox to\wd0{\hss\vrule width4pt height.4pt\hss}%
         \kern-\wd0\copy0\kern-\wd0\raise.35ex\box0}}
\def\Down{{\setbox0\hbox{$\downarrow$}%
         \raise\ht0\hbox to\wd0{\hss\vrule width4pt depth.4pt\hss}%
         \kern-\wd0\box0}}

\newif\ifnadm

\def\doadm{\mathrel{%
   \setbox0 \hbox{$\mathop\vdash$}\dimen0 \ht0
   \setbox0 \hbox{$\vdash$}\advance\dimen0 -\ht0
   \vrule width.8\fontdimen8 \textfont3 height\ht0 depth\dp0
   \mkern-1mu
   \lower\dimen0 \hbox{$\vcenter{%
      \ifnadm
        \setbox0 \hbox{$\scriptstyle\sim\mathstrut$}%
        \hbox{\hbox to\wd0{\hss$\scriptstyle/$\hss}\kern-\wd0 \box0 }%
      \else
        \hbox{$\scriptstyle\sim\mathstrut$}%
      \fi}$}}}

\def\nrstyle#1#2#3{%
  \setbox0\hbox{$#1\bigcirc$}%
  \vcenter{\hbox to\wd0{\hss$#2#3$\hss}}%
  \kern-\wd0\box0 }

\ifx\slide\undefined
  
\fi

\def\st{\expandafter\hat}

\let\lgc\mathbf

\def\IPC{\lgc{IPC}}

\def\FL#1{\lgc{FL_{#1}}}

\def\subhn{\mathcal N_}
\def\subhp{\mathcal P_}

\mathcode`\*="0203

\mathchardef\mhyphen="2D
\def\cput(#1)#2{\put(#1){\hbox to0pt{\hss#2\hss}}}

\ifx\slide\undefined

\def\noproof{\leavevmode\unskip\bme\vadjust{}\nobreak\hfill$\qed$\par}
\let\qed\Box
\newenvironment{Pf}[1][]
  {\par\noindent\textit{Proof\optpar{#1}:}\bme\ignorespaces}
  {\noproof\pagebreak[2]\vskip\medskipamount\ignorespacesafterend}
\def\optpar#1{\ifx\relax#1\relax\else\ #1\fi}
\def\qedhere{\relax\ifmmode\eqno\qed\expandafter\aftergroup
                   \else\noproof\fi\noqed}
\def\noqed{\let\noproof\relax}

\theoremstyle{plain}
\ifx\shortthm\undefined
\newtheorem{Thm}{Theorem}[section]
\else
\newtheorem{Thm}{Theorem}
\fi

\newtheorem{Lem}[Thm]{Lemma}

\newtheorem{Prob}[Thm]{Problem}

\ifx\shortthm\undefined
\def\theCl{\arabic{Cl}}
\fi

\theorembodyfont\upshape
\newtheorem{Def}[Thm]{Definition}
\newtheorem{Rem}[Thm]{Remark}
\newtheorem{Exm}[Thm]{Example}

\newenvironment{Pf*}{\let\qed\qedCl\Pf}\endPf

\fi

\def\theequation{\arabic{equation}}
\def\numberthis{\addtocounter{equation}{1}\tag{\theequation}}

\ifx\slide\undefined
  \usepackage[reftex]{theoremref}
\fi

\newif\iflinenumbers
\linenumberstrue

{\catcode`\^^I=13 \catcode`\^^M=13
\gdef\doalgo#1#2\end#{\hbox to\hsize{\hss \let^^I\qquad%
  \def\\^^M{\nobreak\hfil\break\vadjust{}\qquad}%
  \fboxsep1em \linenum0 %
  \fbox{\hsize#1\vbox{%
  \everypar{\advance\linenum1 %
      \hbox to1.2em{%
           \hss\iflinenumbers$\scriptstyle\the\linenum$\hskip.6em\fi}}%
  #2}}\hss}\end}}
\newcount\linenum

\def\key{\relax\ifmmode\expandafter\mathbf\else\expandafter\textbf\fi}

\def\allowhyphens{\nobreak\hskip0pt\relax}

\DeclareRobustCommand*\magiclparen{\ifmmode(\else\textup(\allowhyphens\fi}
\DeclareRobustCommand*\magicrparen{\ifmmode)\else\textup)\fi}
\let\lparen=(  \let\rparen=)
\def\magicparon{\catcode`\(\active\catcode`\)\active}
\def\magicparoff{\catcode`\(12 \catcode`\)12 }
\AtBeginDocument{\ifx\ifPreview\iftrue\else\magicparon\fi}
\magicparon
\let (=\magiclparen  \let )=\magicrparen

\ifx\sameenum\undefined
  
  \ifx\enumup\undefined
    
  \else
    
  \fi

\else

\fi

\magicparoff

\mathchardef\comma=\mathcode`\,
{\catcode`\,=\active \gdef,{\comma\penalty\relpenalty}}

\ifx\slide\undefined

\providecommand\dedic{\thanks{Supported by
grant IAA100190902 of GA AV \v CR, Center of Excellence CE-ITI under the grant
P202/12/G061 of GA \v CR, and RVO: 67985840.}}

\author{Emil Je\v r\'abek\dedic\\[\medskipamount]
Institute of Mathematics of the Czech Academy of Sciences\\
\small \v Zitn\'a 25,
115\:67 Praha 1,
Czech Republic,
email: \texttt{jerabek@math.cas.cz}
}

\else\ifx

\author{Emil Je\v r\'abek}
\email{jerabek@math.cas.cz\\[-1em]http://math.cas.cz/\string~jerabek/}
\institution{Institute of Mathematics of the Czech Academy of Sciences, Prague}

\else 

\author[Emil Je\v r\'abek]{Emil Je\v r\'abek\\[\medskipamount]
   \scriptsize\texttt{jerabek@math.cas.cz}\\\texttt{http://math.cas.cz/\string~jerabek/}}
\institute{Institute of Mathematics of the Czech Academy of Sciences, Prague}

\fi\fi

\def\pcrl{\mathcal{PCRL}}

\let\dedic\dedicerc

\title{A note on the substructural hierarchy}

\begin{document}
\maketitle

\begin{abstract}
We prove that all axiomatic extensions of the full Lambek calculus with exchange can be axiomatized by
formulas on the $\subhn3$~level of the substructural hierarchy.
\end{abstract}

\section{Introduction}

A standard technique for reduction of the complexity of propositional formulas (nesting depth of connectives) in proof
complexity and other branches of logic is to introduce \emph{extension variables}: we name each subformula by a new
propositional variable, and include appropriate clauses forcing the variables to be equivalent to the original
formulas. This idea may have been independently discovered multiple times; in the context of classical logic,
extension variables appear in the work of Tseitin~\cite{tseit}. Extension variables are systematically used by
Rybakov~\cite{ryb:bk} for the purpose of reducing the formula complexity of nonclassical consequence relations.
It may not be immediately obvious that the method also applies to axioms of substructural logics without contraction,
but as we will see, this can be done with just a little care.

The context we are specifically interested in is the
\emph{substructural hierarchy} introduced by Ciabattoni, Galatos, and Terui~\cite{cia-gal-ter:lics,cia-gal-ter:apal},
stratifying formulas of the full Lambek calculus ($\FL{}$) into classes $\subhp k$ and~$\subhn k$, $k\in\omega$, based
on alternation of polarities of connectives. As shown
in~\cite{cia-gal-ter:lics,cia-gal-ter:apal}, $\subhn2$-axiomatized extensions of $\FL{}$ can be equivalently expressed
by structural rules in the sequent calculus, and similarly,
$\subhp3$~axioms (with certain restrictions) can be expressed by structural hypersequent rules; moreover, analyticity
(subformula property) of the resulting calculi can be characterized algebraically by closure under a certain kind of completion.

We are going to prove
that---at least when the base logic is commutative ($\FL e$)---all remaining axiomatic extensions already appear at the
lowest level of the hierarchy not covered by their results, namely~$\subhn3$. 

\section{Preliminaries}

We refer the reader to Galatos et~al.~\cite{reslat} for comprehensive information on $\FL{}$ and its extensions,
however we include a few words below to clarify our terminology and notation.

The language of $\FL e$ consists of propositional formulas generated from a countable set of variables $p_0,p_1,\dots$
using the connectives $\to,\cdot,\land,\lor,0,1$. We might also include the lattice constants $\bot,\top$; none of our
results depend on their presence or absence (actually, our arguments only rely on the availability of $\to,\land,1$,
and $\to$~alone suffices over~$\FL{ei}$). We abbreviate
\begin{align*}
(\fii\eq\psi)&=(\fii\to\psi)\land(\psi\to\fii),\\
\prod_{i<n}\fii_i&=\fii_0\cdot\fii_1\cdots\fii_{n-1},\\
\fii^n&=\prod_{i<n}\fii,
\end{align*}
with the understanding that the empty product is~$1$.
We write $\psi\sset\fii$ if $\psi$ is a subformula of~$\fii$;
usually, we will need to count multiple occurrences of $\psi$ in~$\fii$ as distinct subformulas.

We employ the notational convention that $\to$ and~$\eq$ bind weaker than other connectives, so that for instance,
\[\prod_{i<3}\fii_i\to\psi=\fii_0\cdot\fii_1\cdot\fii_2\to\psi=(\fii_0\cdot\fii_1\cdot\fii_2)\to\psi.\]

The logic~$\FL e$ can be naturally presented by a sequent calculus, but it will be more convenient for our purposes to
define it using a Hilbert-style calculus: it is axiomatized by a handful of axiom schemata listed in
\cite[Fig.~2.9]{reslat}, and the two rules
\begin{align}
\label{eq:mp}\fii,\fii\to\psi&\ru\psi,\\
\label{eq:adj}\fii&\ru\fii\land1.
\end{align}
If $X$ is a set of formulas, $\FL e+X$ denotes the extension of~$\FL e$ with substitution instances of formulas
from~$X$ as additional axioms. If $L=\FL e+X$, and $\Gamma\cup\{\fii\}$ is a set of formulas, we write
$\Gamma\vdash_L\fii$ if $\fii$ has a derivation in the calculus of~$L$ from a set of premises included in~$\Gamma$. We
will identify $L$ with its consequence relation~$\vdash_L$. Logics of the form $\FL e+X$ are called \emph{axiomatic
extensions} of~$\FL e$. (In general, an \emph{extension} of~$\FL e$ is a Tarski-style consequence relation that contains
$\vdash_{\FL e}$ and is closed under substitution. However, we are not interested in non-axiomatic extensions in this
paper.)

Let $\pcrl$ denote the variety of \emph{pointed commutative residuated lattices}: i.e., structures
$\p{L,{\to},{\cdot},1,{\land},{\lor},0}$ such that $\p{L,{\cdot},1}$ is a commutative monoid, $\p{L,{\land},{\lor}}$ is
a lattice, and
\[x\le y\to z\iff x\cdot y\le z\]
for all $x,y,z\in L$, where $x\le y$ denotes the lattice order $x\land y=x$.

The logic $\FL e$ is \emph{algebraizable} wrt~$\pcrl$:
\begin{gather*}
\fii_1,\dots,\fii_n\vdash_{\FL e}\fii_0
\iff1\land\fii_1\approx1,\dots,1\land\fii_n\approx1\models_\pcrl1\land\fii_0\approx1,\\
\fii_1\approx\psi_1,\dots,\fii_n\approx\psi_n\models_\pcrl\fii_0\approx\psi_0
\iff\fii_1\eq\psi_1,\dots,\fii_n\eq\psi_n\vdash_{\FL e}\fii_0\eq\psi_0,\\
1\land\fii\eq1
\vdash_{\FL e}\fii,\\
1\land(\fii\eq\psi)\approx1
\models_\pcrl\fii\approx\psi.
\end{gather*}
In particular, $\FL e$ is an \emph{equivalential} logic with equivalence connective~$\eq$:
\begin{Lem}\th\label{fact:equi}
For any formulas $\fii,\psi,\chi,\fii',\psi'$, we have
\begin{align}
&\vdash_{\FL e}\fii\eq\fii,\\
\label{eq:7}\fii\eq\psi,\fii\eq\chi&\vdash_{\FL e}\psi\eq\chi,\\
\fii,\fii\eq\psi&\vdash_{\FL e}\psi,\\
\label{eq:5}\fii\eq\fii',\psi\eq\psi'&\vdash_{\FL e}(\fii\circ\psi)\eq(\fii'\circ\psi')
\end{align}
where $\circ\in\{\to,\cdot,\land,\lor\}$.
\noproof\end{Lem}

We will also use the \emph{local deduction theorem} for~$\FL e$ \cite[Cor.~2.15]{reslat}. We include a short proof for
convenience.
\begin{Lem}\th\label{lem:ded}
Let $L$ be an axiomatic extension of~$\FL e$. If $\Gamma,\fii\vdash_L\psi$, then
\begin{equation}\label{eq:3}
\Gamma\vdash_L(\fii\land1)^n\to\psi
\end{equation}
for some $n\in\omega$. If $\pi$ is a (tree-like) $L$-derivation of $\psi$ from $\Gamma\cup\{\fii\}$, we may take for $n$
the number of times the premise $\fii$ is used in~$\pi$.
\end{Lem}
\begin{Pf}
By induction on the length of~$\pi$. If $\psi$ is an axiom of~$L$, or $\psi\in\Gamma$, we can derive $1\to\psi$
from~$\psi$, and $1=(\fii\land1)^0$ by definition. If $\psi=\fii$, we have $\vdash_L\fii\land1\to\fii$.

If $\psi$ is derived from $\chi$ and $\chi\to\psi$ by~\eqref{eq:mp}, we have
\begin{align*}
\numberthis\label{eq:2}
\Gamma&\vdash_L(\fii\land1)^n\to\chi,\\
\Gamma&\vdash_L(\fii\land1)^m\to(\chi\to\psi)\\
\intertext{by the induction hypothesis, which implies}
\Gamma&\vdash_L(\fii\land1)^{n+m}\to\psi
\end{align*}
using
\begin{align*}
&\vdash_{\FL e}\alpha\cdot(\alpha\to\beta)\to\beta,\\
\numberthis\label{eq:1}
\alpha\to\beta,\alpha'\to\beta'&\vdash_{\FL e}\alpha\cdot\beta\to\alpha'\cdot\beta'.
\end{align*}

Finally, if $\psi=\chi\land1$ is derived from~$\chi$ by~\eqref{eq:adj}, the induction hypothesis gives~\eqref{eq:2}.
Using $\vdash_{\FL e}\fii\land1\to1$, $\vdash_{\FL e}1^n\to1$, and~\eqref{eq:1}, we also have
\[\vdash_{\FL e}(\fii\land1)^n\to1,\]
which together with~\eqref{eq:2} yields $\Gamma\vdash_L(\fii\land1)^n\to\chi\land1$.
\end{Pf}

\section{Substructural hierarchy}

The substructural hierarchy introduced in \cite{cia-gal-ter:lics,cia-gal-ter:apal} consists of sets of formulas $\subhp
k$ and~$\subhn k$ for $k\in\omega$, generated by the closure conditions below.
\begin{Def}\th\label{def:subh}
$\subhp k$ and~$\subhn k$ are the smallest sets of formulas with the following properties:
\begin{itemize}
\item $\subhp0=\subhn0$ is the set of propositional variables.
\item $\subhp k\cup\subhn k\sset\subhp{k+1}\cap\subhn{k+1}$.
\item If $\fii,\psi\in\subhp{k+1}$, then $\fii\cdot\psi$, $\fii\lor\psi$, $1$, and~$\bot$ are also in $\subhp{k+1}$.
\item If $\fii,\psi\in\subhn{k+1}$, then $\fii\land\psi$, $0$, and $\top$ are also in $\subhn{k+1}$.
\item If $\fii\in\subhp{k+1}$ and $\psi\in\subhn{k+1}$, then $\fii\to\psi$ is in $\subhn{k+1}$.
\end{itemize}
\end{Def}
The two groups of connectives\footnote{Following a terminology from linear logic,
\cite{cia-gal-ter:lics,cia-gal-ter:apal} call these the positive and negative connectives, respectively, which is what
the letters $\mathcal P$ and~$\mathcal N$ stand for. We avoid these terms here for danger of confusion with
the conventional notion of positive and negative occurrences of subformulas (\th\ref{def:posneg}).} implicit in the definition arise from the
sequent calculus formulation of~$\FL e$: the
left introduction rules for $\cdot,\lor,1,\bot$, and the right introduction rules for $\to,\land,0,\top$, are
invertible.

Our main result shows that for the purpose of classification of axioms over~$\FL e$, the hierarchy collapses
to~$\subhn3$.
\begin{Thm}\th\label{thm:n3}
Every axiomatic extension of~$\FL e$ is axiomatizable by $\subhn3$~formulas.
\end{Thm}
\begin{Pf}
Fix an axiom~$\fii$; we will construct an $\subhn3$~formula $\fii'$ such that $\FL e+\fii=\FL e+\fii'$.

For each occurrence of a subformula $\psi\sset\fii$, we consider a fresh propositional variable~$p_\psi$, and an
associated extension axiom
\[
E_\psi=\begin{cases}
  p_\psi\eq\psi&\text{if $\psi$ is a variable or a constant,}\\
  p_\psi\eq(p_{\psi_0}\circ p_{\psi_1})&\text{if $\psi=\psi_0\circ\psi_1$, $\circ\in\{\to,\cdot,\land,\lor\}$.}
  \end{cases}
\]
Notice that being an equivalence between a variable and a $\subhp1$ or~$\subhn1$ formula, $E_\psi\in\subhn2$.

First, we claim that
\begin{equation}\label{eq:4}
\{E_\chi:\chi\sset\psi\}\vdash_{\FL e}p_\psi\eq\psi,\qquad\psi\sset\fii.
\end{equation}
We prove this by induction on the complexity of~$\psi$. If $\psi$ is a variable or a constant, the right-hand side
of~\eqref{eq:4} is just~$E_\psi$. If $\psi=\psi_0\circ\psi_1$, we have
\[\{E_\chi:\chi\sset\psi\}\vdash_{\FL e}p_{\psi_0}\eq\psi_0,p_{\psi_1}\eq\psi_1
  \vdash_{\FL e}(p_{\psi_0}\circ p_{\psi_1})\eq\psi\]
by the induction hypothesis and~\eqref{eq:5}, hence
\[\{E_\chi:\chi\sset\psi\}\vdash_{\FL e}p_\psi\eq\psi\]
using \eqref{eq:7} and the definition of~$E_\psi$.

Taking $\psi=\fii$ in~\eqref{eq:4}, we obtain
\[\{E_\psi:\psi\sset\fii\}\vdash_{\FL e}\fii\to p_\fii.\]
By the deduction theorem (\th\ref{lem:ded}), we can fix $n\in\omega$ such that
\begin{equation}\label{eq:9}
\vdash_{\FL e}\prod_{\psi\sset\fii}(E_\psi\land1)^n\to(\fii\to p_\fii).
\end{equation}

Let us now define
\[\fii'=\prod_{\psi\sset\fii}(E_\psi\land1)^n\to p_\fii.\]
Since $E_\psi\land1$ is~$\subhn2$, the product is~$\subhp3$, and $\fii'\in\subhn3$ as required. 

We can rewrite \eqref{eq:9} as $\vdash_{\FL e}\fii\to\fii'$, and a fortiori $\vdash_{\FL e+\fii}\fii'$. On the
other hand, let $\sigma$ denote the substitution
\[\sigma(p_\psi)=\psi.\]
Since $\sigma(E_\psi)=(\psi\eq\psi)$ is provable in~$\FL e$, we have
\[\vdash_{\FL e}\sigma\Bigl(\prod_{\psi\sset\fii}(E_\psi\land1)^n\Bigr)\eq\prod_{\psi\sset\fii}1^n\]
which is equivalent to~$1$, thus $\sigma(\fii')$ is equivalent to $\sigma(p_\fii)$, i.e.,
\[\vdash_{\FL e}\sigma(\fii')\to\fii.\]
This gives $\vdash_{\FL e+\fii'}\fii$, hence $\FL e+\fii=\FL e+\fii'$.
\end{Pf}

\begin{Rem}
Let us stress that we restrict attention to \emph{axiomatic} extensions of the base logic because that is the hard case;
axiomatization of general extensions by \emph{rules} of bounded complexity is straightforward. Indeed, it is easy to
see that an arbitrary logic~$L$ (i.e., a structural consequence relation) extending~$\FL{}$ is axiomatized over $\FL{}$ by
rules of the form $\Gamma\ru p$ whose conclusion is a variable, and each formula in~$\Gamma$ is either a variable, or
an equivalence between a variable and a formula containing only one connective; if $L$ is finitary, $\Gamma$ can be
taken finite. The same holds for any (finitely) equivalential base logic in place of~$\FL{}$. In terms of the
substructural hierarchy, this means that all extensions of~$\FL{}$ are axiomatizable by rules with $\subhn2$~premises, and $\subhn0$~conclusions.
\end{Rem}

A concrete illustration of \th\ref{thm:n3} is given later in \th\ref{exm:cint}. We have to postpone it for
the following reason: the $\subhn3$ axiom~$\fii'$ constructed in the proof of \th\ref{thm:n3} is not presented fully
explicitly, as it depends
on~$n$. We can in principle compute~$n$ for a given~$\fii$ as the proofs of \th\ref{fact:equi,lem:ded} are
constructive, but in fact, we can do better: digging a bit deeper into the guts of the argument will reveal that we can
just take $n=1$ for all~$\fii$; moreover, we can shorten~$\fii'$ somewhat by employing implications instead of equivalences,
distinguishing between positively and negatively occurring subformulas of~$\fii$. We now present the details.

First, let us recall the concept of positive and negative occurrences.
\begin{Def}\th\label{def:posneg}
An occurrence of a subformula~$\psi$ in~$\fii$ is classified as \emph{positive} or \emph{negative} as follows.
\begin{itemize}
\item The occurrence of $\fii$ in itself is positive.
\item For any positive (negative) occurrence of $\psi_0\circ\psi_1$ in~$\fii$, where
$\circ\in\{\cdot,\land,\lor\}$, the indicated occurrences of $\psi_0$ and~$\psi_1$ in~$\fii$ are also positive
(negative, resp.).
\item For any positive (negative) occurrence of $\psi_0\to\psi_1$ in~$\fii$, the indicated occurrence of $\psi_0$
in~$\fii$ is negative (positive, resp.), and the occurrence of~$\psi_1$ is positive (negative, resp.).
\end{itemize}
Let us abbreviate
\[(\fii\TO\psi)=(\fii\to\psi)\land1.\]
\end{Def}
\begin{Lem}\th\label{lem:mon}
$\FL e$ proves the schemata
\begin{gather}
\fii\TO\fii,\\
\label{eq:24}(\fii\TO\psi)\cdot(\psi\TO\chi)\to(\fii\TO\chi),\\
\label{eq:21}(\fii'\TO\fii)\cdot(\psi\TO\psi')\to\bigl((\fii\to\psi)\TO(\fii'\to\psi')\bigr),\\
\label{eq:25}(\fii\TO\fii')\cdot(\psi\TO\psi')\to\bigl((\fii\circ\psi)\TO(\fii'\circ\psi')\bigr)
\end{gather}
for $\circ\in\{\cdot,\land,\lor\}$.
\end{Lem}
\begin{Pf}
Straightforward, using e.g.\ the algebraic semantics of~$\FL e$.

For instance, let us check \eqref{eq:25} with $\circ=\land$. Let $L$ be a residuated lattice, and $x,x',y,y'\in L$, we
need to show
\[(x\TO x')\cdot(y\TO y')\le(x\land y)\TO(x'\land y').\]
Clearly,
\[(x\TO x')\cdot(y\TO y')\le1\cdot1=1,\]
thus it suffices to show
\[(x\TO x')\cdot(y\TO y')\le(x\land y)\to(x'\land y').\]
This follows from
\[(x\land y)\cdot(x\TO x')\cdot(y\TO y')\le x\cdot(x\to x')\cdot1\le x',\]
and the symmetric inequality for~$y'$.
\end{Pf}
\begin{Def}\th\label{def:montr}
Let $\fii$ be a formula. We will define an $\subhn3$~formula $\fii^+$ as follows.

If $\psi$ is an occurrence of a variable\footnote{The separate treatment of
variables only serves the purpose of making $\fii^+$ shorter, otherwise we could handle them more uniformly as in the
proof of \th\ref{thm:n3}.} in~$\fii$, we consider $p_\psi$ a shorthand for~$\psi$. For any occurrence of a subformula
$\psi\sset\fii$ which is not a variable, we introduce a new variable~$p_\psi$, and put
\[
E^+_\psi=\begin{cases}
  \psi\TO p_\psi&\text{if $\psi$ is a constant and occurs positively in~$\fii$,}\\
  p_\psi\TO\psi&\text{if $\psi$ is a constant and occurs negatively in~$\fii$,}\\
  (p_{\psi_0}\circ p_{\psi_1})\TO p_\psi&\text{if $\psi=\psi_0\circ\psi_1$ occurs positively in~$\fii$,}\\
  p_\psi\TO(p_{\psi_0}\circ p_{\psi_1})&\text{if $\psi=\psi_0\circ\psi_1$ occurs negatively in~$\fii$,}
  \end{cases}
\]
where $\circ\in\{\to,\cdot,\land,\lor\}$. Finally,
\[\fii^+=\prod_{\psi\sset'\fii}E^+_\psi\to p_\fii,\]
where $\psi\sset'\fii$ means that $\psi\sset\fii$ and $\psi$ is not a variable.

We observe that $E^+_\psi$ is~$\subhn2$, and $\fii^+$ is~$\subhn3$.
\end{Def}
\begin{Rem}\th\label{rem:occ}
When there are multiple occurrences of the same formula $\psi$ in~$\fii$, each gets its own variable~$p_\psi$ according
to the given definition. This is not really essential, but what matters is that $\fii^+$ includes one~$E^+_\psi$ for every occurrence.
\end{Rem}
\begin{Thm}\th\label{thm:n3-mon}
For any formula~$\fii$, the $\subhn3$~formula $\fii^+$ satisfies $\FL e+\fii=\FL e+\fii^+$.

More precisely, $\FL e$ proves
\begin{gather}
\label{eq:18}\fii\to\fii^+,\\
\label{eq:16}\sigma(\fii^+)\to\fii,
\end{gather}
where $\sigma$ denotes the substitution $\sigma(p_\psi)=\psi$.
\end{Thm}
\begin{Pf}
The same argument as in the proof of \th\ref{thm:n3} shows~\eqref{eq:16}.

As for~\eqref{eq:18}, we prove by induction on the complexity of $\psi\sset\fii$ that
\begin{equation}\label{eq:17}
\vdash_{\FL e}\prod_{\chi\sset'\psi}E^+_\chi\to(\psi\TO p_\psi)
\end{equation}
if the occurrence of $\psi$ in~$\fii$ is positive, and
\begin{equation}
\vdash_{\FL e}\prod_{\chi\sset'\psi}E^+_\chi\to(p_\psi\TO\psi)
\end{equation}
if it is negative.

The claim is immediate from the definition if $\psi$ is a constant or a variable.

Let $\psi=\psi_0\to\psi_1$. If $\psi$ occurs positively, we have
\[\prod_{\chi\sset'\psi}E^+_\chi
=\biggl(\prod_{\chi\sset'\psi_0}E^+_\chi\biggr)
\cdot\biggl(\prod_{\chi\sset'\psi_1}E^+_\chi\biggr)
\cdot\bigl((p_{\psi_0}\to p_{\psi_1})\TO p_\psi\bigr),\]
and the induction hypothesis gives
\begin{align*}
&\vdash_{\FL e}\prod_{\chi\sset'\psi_0}E^+_\chi\to(p_{\psi_0}\TO\psi_0),\\
&\vdash_{\FL e}\prod_{\chi\sset'\psi_1}E^+_\chi\to(\psi_1\TO p_{\psi_1}),
\end{align*}
thus
\[\vdash_{\FL e}\prod_{\chi\sset'\psi}E^+_\chi
  \to(p_{\psi_0}\TO\psi_0)\cdot(\psi_1\TO p_{\psi_1})\cdot\bigl((p_{\psi_0}\to p_{\psi_1})\TO p_\psi\bigr).\]
Using~\eqref{eq:21}, this implies
\[\vdash_{\FL e}\prod_{\chi\sset'\psi}E^+_\chi
  \to\bigl((\psi_0\to\psi_1)\TO(p_{\psi_0}\to p_{\psi_1})\bigr)\cdot\bigl((p_{\psi_0}\to p_{\psi_1})\TO p_\psi\bigr),\]
hence
\[\vdash_{\FL e}\prod_{\chi\sset'\psi}E^+_\chi\to\bigl((\psi_0\to\psi_1)\TO p_\psi\bigr)\]
by~\eqref{eq:24}.

If $\psi$ occurs negatively in~$\fii$, the induction hypothesis and the definition of~$E^+_\psi$ give
\[\vdash_{\FL e}\prod_{\chi\sset'\psi}E^+_\chi
  \to(\psi_0\TO p_{\psi_0})\cdot(p_{\psi_1}\TO\psi_1)\cdot\bigl(p_\psi\TO(p_{\psi_0}\to p_{\psi_1})\bigr),\]
which implies
\[\vdash_{\FL e}\prod_{\chi\sset'\psi}E^+_\chi\to\bigl(p_\psi\TO(\psi_0\to\psi_1)\bigr)\]
in a similar way using \eqref{eq:24} and~\eqref{eq:21}.

If $\psi=\psi_0\circ\psi_1$ with $\circ\in\{\cdot,\land,\lor\}$, we proceed analogously with \eqref{eq:25} in place
of~\eqref{eq:21}.

Taking $\psi=\fii$ in~\eqref{eq:17} gives
\[\vdash_{\FL e}\prod_{\psi\sset'\fii}E^+_\psi\to(\fii\to p_\fii)\]
using the definition of~$\TO$, thus $\vdash_{\FL e}\fii\to\fii^+$.
\end{Pf}
\iffalse
\begin{Exm}\th\label{exm:cint}
Let $\fii$ be Cintula's product axiom (cf.~\cite[p.~114]{reslat})
\[\bigl((p\to0)\to0\bigr)\to\bigl[(p\to p\cdot q)\to q\cdot((q\to0)\to0)\bigr],\]
which is ostensibly~$\subhn4$. Then $\fii^+$ is
\begin{align*}
\bigl[&(0\TO z_0)\cdot((p\to z_0)\TO p')\cdot(z_1\TO0)\cdot(p''\TO(p'\to z_1))\cdot(r\TO p\cdot q)\cdot(s\TO(p\to r))\\
&\cdot(z_2\TO0)\cdot(q'\TO(q\to z_2))\cdot(0\TO z_3)\cdot((q'\to z_3)\TO q'')\cdot(q\cdot q''\TO t)\\
&\cdot((s\to t)\TO u)\cdot((p''\to u)\TO v)\bigr]\to v,
\end{align*}
where the extension variables have been renamed to spare obnoxious subscripts: $z_i$ correspond to the occurrences of~$0$; $p'$ to
$p\to0$, $p''$ to $(p\to0)\to0$, and similarly for $q'$, $q''$; $r$ to $p\cdot q$; $s$ to $p\to p\cdot q$; $t$ to
$q\cdot((q\to0)\to0)$; $u$ to the subformula of~$\fii$ in square brackets; and $v$ to the whole formula. According to
\th\ref{rem:occ}, we could have replaced $z_0,z_1,z_2,z_3$ with a single variable~$z$. It turns out that since $\fii$
contains no lattice connectives, it would also suffice to use plain $\to$ rather than~$\TO$.
\end{Exm}
\else
\begin{Exm}\th\label{exm:cint}
Let $\fii$ be Cintula's product axiom (cf.~\cite[p.~114]{reslat})
\[\bigl((r\to0)\to0\bigr)\to\bigl[(r\to r\cdot q)\to q\cdot((q\to0)\to0)\bigr],\]
which is ostensibly~$\subhn4$. Then $\fii^+$ is
\begin{align*}
\begin{aligned}
\bigl[
&(0\TO p_{0,0})\cdot\bigl((r\to p_{0,0})\TO p_{\neg r}\bigr)\cdot(p_{0,1}\TO0)
 \cdot\bigl(p_{\neg\neg r}\TO(p_{\neg r}\to p_{0,1})\bigr)\\
&\cdot(p_{r\cdot q}\TO r\cdot q)\cdot\bigl(p_{r\to r\cdot q}\TO(r\to p_{r\cdot q})\bigr)
 \cdot(p_{0,2}\TO0)\cdot\bigl(p_{\neg q}\TO(q\to p_{0,2})\bigr)\\
&\cdot(0\TO p_{0,3})\cdot\bigl((p_{\neg q}\to p_{0,3})\TO p_{\neg\neg q}\bigr)
 \cdot(q\cdot p_{\neg\neg q}\TO p_{q\cdot\neg\neg q})\\
&\cdot\bigl((p_{r\to r\cdot q}\to p_{q\cdot\neg\neg q})\TO p_{(r\to r\cdot q)\to q\cdot\neg\neg q}\bigr)
 \cdot\bigl((p_{\neg\neg r}\to p_{(r\to r\cdot q)\to q\cdot\neg\neg q})
                    \TO p_{\neg\neg r\to((r\to r\cdot q)\to q\cdot\neg\neg q)}\bigr)\bigr]
\end{aligned}&\\
\to p_{\neg\neg r\to((r\to r\cdot q)\to q\cdot\neg\neg q)}&,
\end{align*}
where we used the abbreviation $\neg\alpha=(\alpha\to0)$ in the subscripts, and the four extension variables corresponding to occurrences of~$0$ were disambiguated by extra subscripts
$0,\dots,3$. We could have actually used just a single variable~$p_0$, cf.\ \th\ref{rem:occ}. It turns out that
since $\fii$ contains no lattice connectives, it would also suffice to use plain $\to$ rather than~$\TO$.

In contrast, the corresponding formula~$\fii'$ from \th\ref{thm:n3} is
\begin{align*}
\begin{aligned}
\bigl[
&(p_{r,0}\EQ r)\cdot(p_{0,0}\EQ0)\cdot\bigl(p_{\neg r}\EQ(p_{r,0}\to p_{0,0})\bigr)\cdot(p_{0,1}\EQ0)
 \cdot\bigl(p_{\neg\neg r}\EQ(p_{\neg r}\to p_{0,1})\bigr)\\
&\cdot(p_{r,2}\EQ r)\cdot(p_{q,0}\EQ q)\cdot(p_{r\cdot q}\EQ p_{r,2}\cdot p_{q,0})
 \cdot(p_{r,1}\EQ r)\cdot\bigl(p_{r\to r\cdot q}\EQ(p_{r,1}\to p_{r\cdot q})\bigr)\\
&\cdot(p_{q,2}\EQ q)\cdot(p_{0,2}\EQ0)\cdot\bigl(p_{\neg q}\EQ(p_{q,2}\to p_{0,2})\bigr)\\
&\cdot(p_{0,3}\EQ0)\cdot\bigl(p_{\neg\neg q}\EQ(p_{\neg q}\to p_{0,3})\bigr)
 \cdot(p_{q,1}\EQ q)\cdot(p_{q\cdot\neg\neg q}\EQ p_{q,1}\cdot p_{\neg\neg q})\\
&\cdot\bigl(p_{(r\to r\cdot q)\to q\cdot\neg\neg q}\EQ(p_{r\to r\cdot q}\to p_{q\cdot\neg\neg q})\bigr)
 \cdot\bigl(p_{\neg\neg r\to((r\to r\cdot q)\to q\cdot\neg\neg q)}
     \EQ(p_{\neg\neg r}\to p_{(r\to r\cdot q)\to q\cdot\neg\neg q})\bigr)\bigr]
\end{aligned}&\\
\to p_{\neg\neg r\to((r\to r\cdot q)\to q\cdot\neg\neg q)}&,
\end{align*}
where $\alpha\EQ\beta$ stands for $(\alpha\eq\beta)\land1$. Here we use the fact that we can take $n=1$
in~\eqref{eq:9}, which can be proved in a similar way as \th\ref{thm:n3-mon}.
\end{Exm}
\fi

For ease of reference in the next remark, we state a normal form for $\subhp k$ and~$\subhn k$ formulas proved in
\cite[Lemma~3.3]{cia-gal-ter:apal}. Recall that the empty product is~$1$; likewise, empty disjunctions and (lattice)
conjunctions are defined as $\bot$ and~$\top$, respectively.
\begin{Lem}\th\label{lem:nf}
Let $k\ge0$.
\begin{enumerate}
\item\label{item:1}
Any $\subhp{k+1}$~formula is equivalent over~$\FL e$ to a disjunction of products of $\subhn k$~formulas.
\item\label{item:2}
Any $\subhn{k+1}$~formula is equivalent over~$\FL e$ to $\ET_{i<n}(\alpha_i\to\beta_i)$, where each $\alpha_i$ is a
product of $\subhn k$~formulas, and each $\beta_i$ is a $\subhp k$~formula or~$0$.
\noproof\end{enumerate}
\end{Lem}

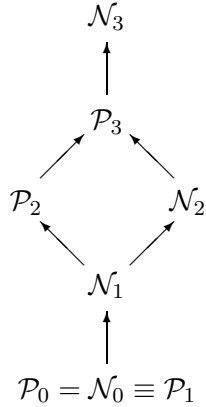
\begin{figure}
\centering
\magicparoff
\unitlength=1em
\begin{picture}(8,13.5)
\cput(4,0){$\subhp0=\subhn0\equiv\subhp1$}
\put(4,1.2){\vector(0,1){1.8}}
\cput(4,3.6){$\subhn1$}
\put(3.2,4.6){\vector(-1,1){1.5}}
\put(4.8,4.6){\vector(1,1){1.5}}
\cput(1.2,6.5){$\subhp2$}
\cput(6.8,6.5){$\subhn2$}
\put(1.7,7.6){\vector(1,1){1.5}}
\put(6.3,7.6){\vector(-1,1){1.5}}
\cput(4,9.3){$\subhp3$}
\put(4,10.5){\vector(0,1){1.8}}
\cput(4,12.9){$\subhn3$}
\end{picture}
\caption{The substructural hierarchy}
\label{fig:subh}
\end{figure}
\begin{Rem}\th\label{rem:coll}
Figure~1 shows what is left of the substructural hierarchy over~$\FL e$. Concerning $\subhp1\equiv\subhp0$, any
$\fii(p_1,\dots,p_n)\in\subhp1$ can be written as a disjunction of products of variables by
\th\ref{lem:nf}~\ref{item:1}. If one of the products is empty, $\fii$ is provable in~$\FL e$; otherwise
$\fii(p\land1,\dots,p\land1)$ implies~$p$. Thus, the only $\subhp1$-axiomatizable logics are $\FL e$ itself and the
inconsistent logic.

The hierarchy is not going to collapse any further, as all remaining inclusions are strict:

An example of a nontrivial $\subhn1$~axiom is left weakening $p\to(q\to p)$.

By \cite[Cor.~7.7]{cia-gal-ter:apal}, the $\subhp2$ linearity axiom $(p\to q)\lor(q\to p)$ is not
$\subhn2$-axiomatizable. The same holds for the law of excluded middle $p\lor(p\to0)$.

The right weakening axiom $0\to p$ is~$\subhn2$, but it is not $\subhp2$-axiomatizable over Johansson's logic
($\FL{eci}$). Assuming otherwise, it would be axiomatizable by disjunctions of $\subhn1$~axioms over~$\FL{eci}$
by~\th\ref{lem:nf}~\ref{item:1}, using ${\cdot}={\land}$. Since $\FL{eci}+(0\to p)=\IPC$ has the disjunction
property, we could replace each disjunction with one of its disjuncts, hence the logic would be actually
$\subhn1$-axiomatizable. By~\ref{item:2}, we could axiomatize it by a set of axioms of the form $\alpha\to\beta$,
where $\alpha$ is a product of variables, and $\beta$ is a variable or~$0$. However, such an axiom is valid
in~$\IPC$ only when $\beta$ is a variable occurring in~$\alpha$, in which case it is already provable in~$\FL{eci}$,
hence this is impossible.

Finally, a proper superintuitionistic logic with the disjunction property, such as $\lgc{KP}=\IPC+(\neg p\to q\lor
r)\to(\neg p\to q)\lor(\neg p\to r)$, is not $\subhp3$-axiomatizable over~$\IPC$. Assuming otherwise, the same argument
as above would imply the logic is in fact $\subhn2$-axiomatizable. However, as shown
in~\cite{cia-gal-ter:apal}, any $\subhn2$~axiom is either provable or contradictory over~$\IPC$.
\end{Rem}

\section{Conclusion}

We have seen that over~$\FL e$, arbitrary axioms can be unwinded to deductively equivalent $\subhn3$~axioms, hence the
substructural hierarchy collapses. This entails some ramifications for the program of algebraic proof theory: the
optimist may say that now it suffices to extend the structure theory for $\subhn2$ and~$\subhp3$ logics just one step
to~$\subhn3$ to deal with arbitrary extensions of~$\FL e$, while the pessimist may point out that this
sounds too good to be feasible, and it rather means that the class~$\subhn3$ as a whole is already intractable to
informative analysis, and might need further subclassification.

Our arguments relied on commutativity, which raises the question what happens if we drop this assumption:
\begin{Prob}\th\label{prob:noncom}
Are all axiomatic extensions of~$\FL{}$ $\subhn k$-axiomatizable for some fixed~$k$?
\end{Prob}
We mention that while the basic structure of the proof of \th\ref{thm:n3}---which essentially uses only the equivalentiality
of the logic and the deduction theorem---applies to~$\FL{}$ as well, this does not yield the desired reduction in
formula complexity. The problem is that the form of deduction theorem valid for~$\FL{}$ has $\fii\land1$ in~\eqref{eq:3}
replaced with iterated conjugates $\gamma_1(\gamma_2(\dots(\gamma_m(\fii))\dots))$, where each $\gamma_i(x)$ is
$(\alpha_i\backslash(x\cdot\alpha_i))\land1$ or $((\alpha_i\cdot x)/\alpha_i)\land1$ for some formulas~$\alpha_i$. Even
if we disregard the complexity of~$\alpha_i$ itself (which we can't), each conjugate strictly raises the level in the
substructural hierarchy, hence the resulting formula may have unbounded complexity.

The low-level proof of \th\ref{thm:n3-mon} does not work in the noncommutative setting either. The argument relies on
exchange through repeated use of \th\ref{lem:mon}; it is unclear whether one can choose an ordering of the factors
in the definition of~$\fii^+$ and directions of the relevant residua in a consistent way so that everything cancels out
as intended.

We thus leave \th\ref{prob:noncom} open.

\section*{Acknowledgement}
I am grateful to Agata Ciabattoni and Nick Galatos for persuading me that the results in this paper are not generally
known, and to Agata Ciabattoni for useful comments on a preliminary version of the manuscript.

\bibliographystyle{mybib}
\bibliography{mybib}
\end{document}
